\documentclass[prb,reprint,amsmath,amssymb,showpacs,superscriptaddress,floatfix,longbibliography]{revtex4-1}
\usepackage[breaklinks=true,colorlinks,citecolor=blue,linkcolor=blue,urlcolor=blue]{hyperref}

\usepackage{epsfig,mathrsfs,color,latexsym,subfigure,marginnote,gensymb,}
\usepackage{graphicx}

\renewcommand{\BibitemShut}[1]{}

\begin{document}
\title{Interlayer decoupling in twisted bilayers of $\beta$-phosphorus and arsenic: a computational study}
\author{Shantanu Agnihotri}
\affiliation{Department of Electrical Engineering, Indian Institute of Technology Kanpur, Kanpur, U.P., 208016, India}
\author{Maneesh Kumar}
\affiliation{Department of Materials Science and Engineering, Indian Institute of Technology Kanpur, Kanpur, U.P., 208016, India}
\author{Yogesh Singh Chauhan}
\affiliation{Department of Electrical Engineering, Indian Institute of Technology Kanpur, Kanpur, U.P., 208016, India}
\author{Amit Agarwal}
\affiliation{Department of Physics, Indian Institute of Technology Kanpur, Kanpur, U.P., 208016, India}
\author{Somnath Bhowmick}
\email{bsomnath@iitk.ac.in}
\affiliation{Department of Materials Science and Engineering, Indian Institute of Technology Kanpur, Kanpur, U.P., 208016, India}
\date{\today}

\begin{abstract}
We investigate magnetism and band structure engineering in Moir\'e superlattice of blue phosphorus ($\beta$-P) and grey arsenene ($\beta$-As) bilayers, using \textit{ab initio} calculations.  The electronic states near the valence and conduction band edges have significant $p_z$ character in both the bilayers. Thus, twisting the layers significantly reduce the interlayer orbital overlap, leading to a decrease in the binding energy (up to $\sim33\%$) and an increase in interlayer distance (up to $\sim10\%$), compared to the most stable AA-stacking. This interlayer decoupling also results in a notable increase (up to $\sim$25-50\%) of the bandgap of twisted bilayers, with the valance band edge becoming relatively flat with van-Hove singularities in the density of states. Thus, hole doping induces a Stoner instability, leading to  ferromagnetic ground state, which is more robust in Moir\'e superlattices, than that of AA-stacked $\beta$-P and $\beta$-As. 
\end{abstract}

\maketitle
\section{Introduction}
The era of 2D materials started with the discovery of graphene, hosting massless Dirac quasiparticles, with gate tunable electronic and optical properties.\cite{Novoselov,Castro,ZHONG201720} This motivated the discovery of several other 2D semiconductors. Among them, transition metal dichalcogenides (such as MoS$_2$, WS$_2$ etc.)\cite{Chhowalla, Xiaodong,XIA20171} and black phosphorene\cite{Neal,LIN201715} are some of the predominantly explored candidates. Several useful applications including transistors based on graphene,\cite{Schwierz,PAEK201771} MoS$_2$,\cite{Giacometti} and black phosphorene\cite{Yijun} have already been demonstrated.\cite{Zeng} There has also been significant interest in twisted bilayers forming Moir\'e superlattices, with the twist angle offering another handle on the tunability of electronic and optical properties.\cite{mosaic,Kim3364,Magic-angle} In particular, several recent studies on twisted bilayer graphene explore the presence of flat bands and the van-Hove singularities (vHs) resulting in exciting many body instabilities, on account of the relative rotation between the layers.\cite{tan2016, Santos, Andrei, Mallet, Mengxi,padhi2018} 

The properties of ultrathin layered materials are governed by the stacking sequence of the monolayers, which dictates the interlayer orbital overlap among different layers. Thus, understanding the influence of the stacking sequences on the nature of the interlayer interaction among monolayers has been of great fundamental interest.\cite{ping2012,julia2011,Liu2014,yan2015,kecik1,Zhang2015,Shulenburger2015} Motivated by this, in this paper we explore the nature of the interlayer coupling and the modulation of the electronic and magnetic properties in twisted bilayers of blue phosphorus ($\beta$-P)\cite{David,Tom} and grey arsenene ($\beta$-As),\cite{Mardanya,kamal2015} both having similar crystal structure and physical properties. Among several possible 2D allotropes predicted for group-V elements,\cite{Bajaj} $\beta$-P has already been reported experimentally,\cite{zhang2016, gu2017} along with a detailed atomistic study of its growth mechanism.\cite{han2017} $\beta$-allotrope of group-V elements have a graphene like buckled honeycomb structure, as opposed to the puckered honeycomb crystal of black phosphorene or $\alpha$-P. In terms of electronic transport properties, both the allotropes are found to be comparable.\cite{Achintya2018} Interestingly, a hole doping induced ferromagnetic ground state has been reported in case of $\beta$-P.\cite{botao2017,zhou2017}

Our study reveals that the $p_z$ orbitals contribute significantly to the valance as well as conduction band edge of the bilayers of AA-stacked $\beta$-P and $\beta$-As. Thus, the interactions among the $p_z$ orbitals on different layers are impacted significantly by the relative rotation of the layers relative to each other. We find that, increasing the rotation angle leads to larger interlayer distances and reduced binding energy in both $\beta$-As and $\beta$-P bilayers. As a result, the bandgap increases in Moir\'e superlattices, and also shows more modulation with  applied electric field and strain. Additionally, the highest valance band of the Moir\'e superlattices also becomes relatively flat with van Hove singularities in the density of states. This leads to Stoner instability induced ferromagnetic ground state with hole doping, which is found to be more stable in twisted structures, than that of AA-stacked bilayers. 

The manuscript is organized as follows: In Sec.~\ref{sec2}, we describe the details of the density functional theory (DFT) calculations. The electronic properties and the interlayer decoupling of the Moir\'e superlattice are discussed in detail in the Sec.~\ref{sec3}; in the context of binding energy, electronic band structure and its modulation with vertical electric field and strain, and magnetic instability in Moir\'e superlattices. The results are summarized in Sec.~\ref{sec4}.

\begin{figure*}
\centering
\includegraphics[width=\linewidth]{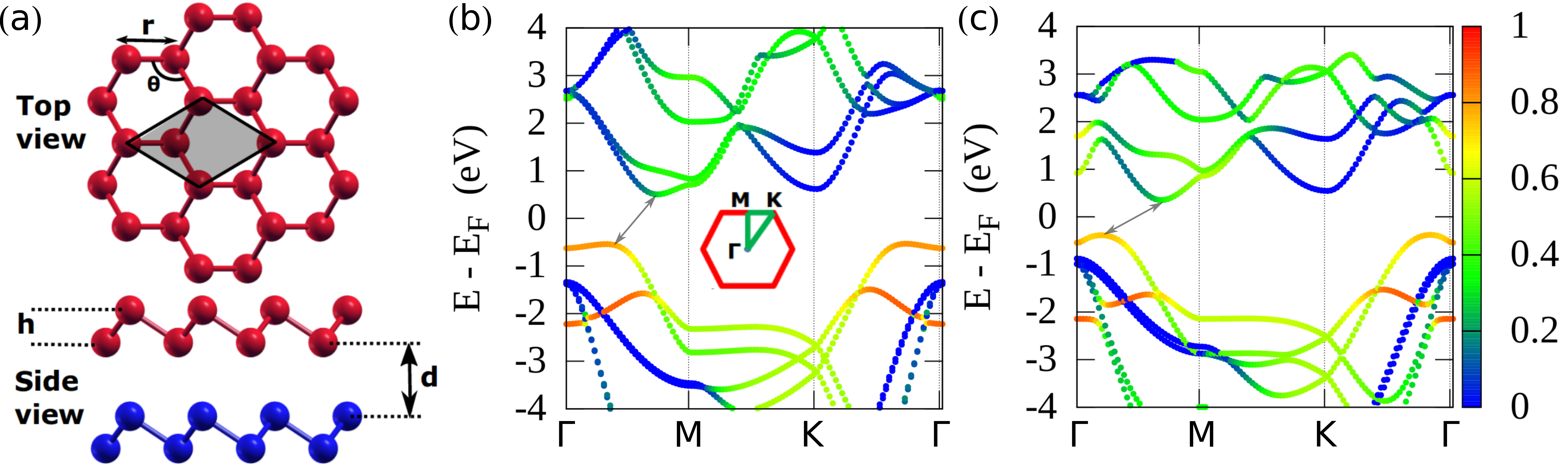}
\caption{(a) The crystal structure of bilayer $\beta$-P and $\beta$-As. On account of the AA-stacking, both the top (red) and bottom (blue) layer are visible separately in the side view only. The unit cell is highlighted in the top view. The Orbital resolved electronic band structures (plotted along the high symmetry lines shown in the inset) of bilayer $\beta$-P and $\beta$-As are illustrated in (b) and (c), respectively. The arrow connects the valence band maximum (VBM) and conduction band minimum (CBM). The color scale shows the relative weight of the $p_z$ orbitals on different energy bands. Clearly, the  $p_z$ orbitals dominate near the VBM, and have significant contribution near the CBM as well.}
\label{fig1}
\end{figure*}

\section{Methodology}
\label{sec2}
Structural relaxations and electronic band structure calculations are performed using density functional theory (DFT), as implemented in the Quantum ESPRESSO package.\cite{Giannozzi} A plane wave basis set with kinetic energy cutoff of 30 Ry and projector augmented wave pseudopotentials are used. Exchange-correlation effects are included within the framework of generalized gradient approximations (GGA), as proposed by Perdrew-Burke-Ernzerhof (PBE).\cite{Ernzerhof} Dispersive forces are taken into account by using the vdW-DF-obk8 van der Waals correction.\cite{Zuluaga, Valentino, Langreth} A $k$-point mesh of $24\times 24\times 1$ is used for Brillouin zone integrations in case of smaller unit cells of AA-stacked bilayers and a coarser mesh is chosen for the Moir\'e superlattices in twisted bilayers, according to their size. Structural optimizations are carried out until the energy difference between two successive steps of ionic relaxation are less than $10^{-4}$ Ry and all three components of the force on each atom belonging to the unit cell are less than $10^{-3}$ Ry/Bohr. A vacuum of 20 \AA~ is applied perpendicular to the plane to avoid spurious interactions between the periodic replicas in the out-of-plane direction. The crystal structures are prepared by XCrysden software.\cite{Kokalj03}

\par

\begin{figure*}
\centering
\includegraphics[width=\linewidth]{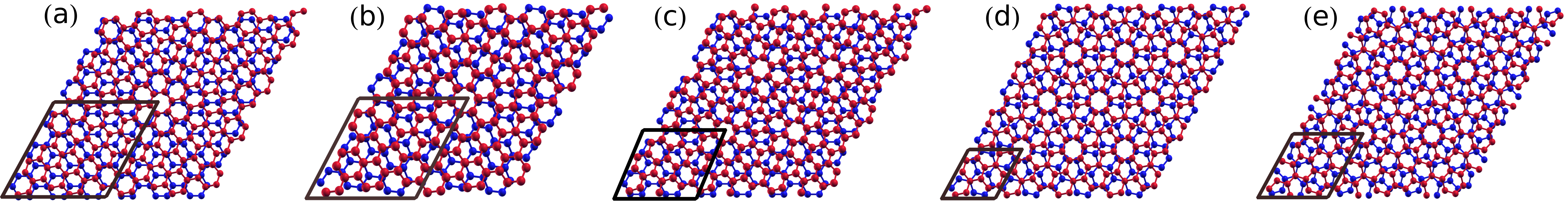}
\caption{Moir\'e superlattices of $\beta$-P and $\beta$-As (with a hexagonal unit cell), as obtained by rotating one layer with respect to the other by five different angles. Rotation angles (number of atoms per unit cell) in different panels are: (a) 10.9$^{\circ}$ (110), (b) 13.2$^{\circ}$ (72), (c) 16.1$^{\circ}$ (50), (d) 21.8$^{\circ}$ (28) and (e) 27.8$^{\circ}$ (52).}
\label{fig2}
\end{figure*}

\section{Results and Discussions}
\label{sec3}

\subsection{AA-stacked bilayers: crystal and electronic band structure}
The $\beta$-allotrope of monolayer P and As are known to have a buckled honeycomb structure.\cite{Barun,kecik1,Mardanya,Bajaj} Among several possible stacking sequences, AA-stacked bi-layers are known to have the lowest energy configuration.\cite{Barun,kecik1} The top and side view of the crystal are shown in Fig.~\ref{fig1}(a). The values of the structural parameters marked in Fig.~\ref{fig1}(a), like the bond length (r), bond angle ($\theta$), buckling height (h) and interlayer separation (d), are reported in Table~\ref{table1}. These are in good agreement with the earlier studies on $\beta$-P \cite{Barun} (reported values of r = 2.26 \AA, $\theta=93.06^\circ$, h = 1.24 \AA, d = 3.23 \AA) and $\beta$-As\cite{kecik1} (reported values of r = 2.51 \AA, $\theta=91.87^\circ$, h = 1.4 \AA, d = 3.27 \AA). 

Surveying the structural data available in the literature for these bilayers, we find that the interlayer distance is sensitive to the choice of van der Waals interaction parameters, and values ranging from 3.23 to 3.40 \AA~ [3.14 to 3.27 \AA] are reported for AA-stacked $\beta$-P [$\beta$-As].\cite{Barun,Pontes2018,kecik1,kecik2} Unlike the inter-layer distance, other structural parameters like unit cell dimension, bond length and bond angle are independent of the choice of the van der Waals interaction, and the reported values lie within a very narrow range.\cite{Barun,Pontes2018,kecik1,kecik2} In our calculations, the interlayer distance lies within the range of values reported in the literature, and the other structural parameters match very closely (within 1-2\%). This validates our choice of the pseudo-potential and other calculation parameters like kinetic energy cut-off and k-point mesh used in this work. 

\begin{table}[t]
\caption{Structural parameters and bandgap of AA-stacked bilayer $\beta$-P and $\beta$-As. Our calculations are in good agreement with earlier works on $\beta$-P\cite{Barun} and $\beta$-As\cite{kecik1} (see main text for details).}
\label{table1}
\begin{tabular}{c c c c c c}
\hline
\hline
Crystal & r (\AA) & $\theta$ & h (\AA) & d (\AA) & E$_g$ (eV)\\
\hline
$\beta$-P & 2.26 & 93.06$^\circ$ & 1.24 & 3.35 & 1.17 \\
\hline
$\beta$-As & 2.51 & 92.14$^\circ$ & 1.39 & 3.2 & 0.86 \\
\hline 
\hline
\end{tabular}
\end{table}

\begin{figure}
\centering
\includegraphics[width=\linewidth]{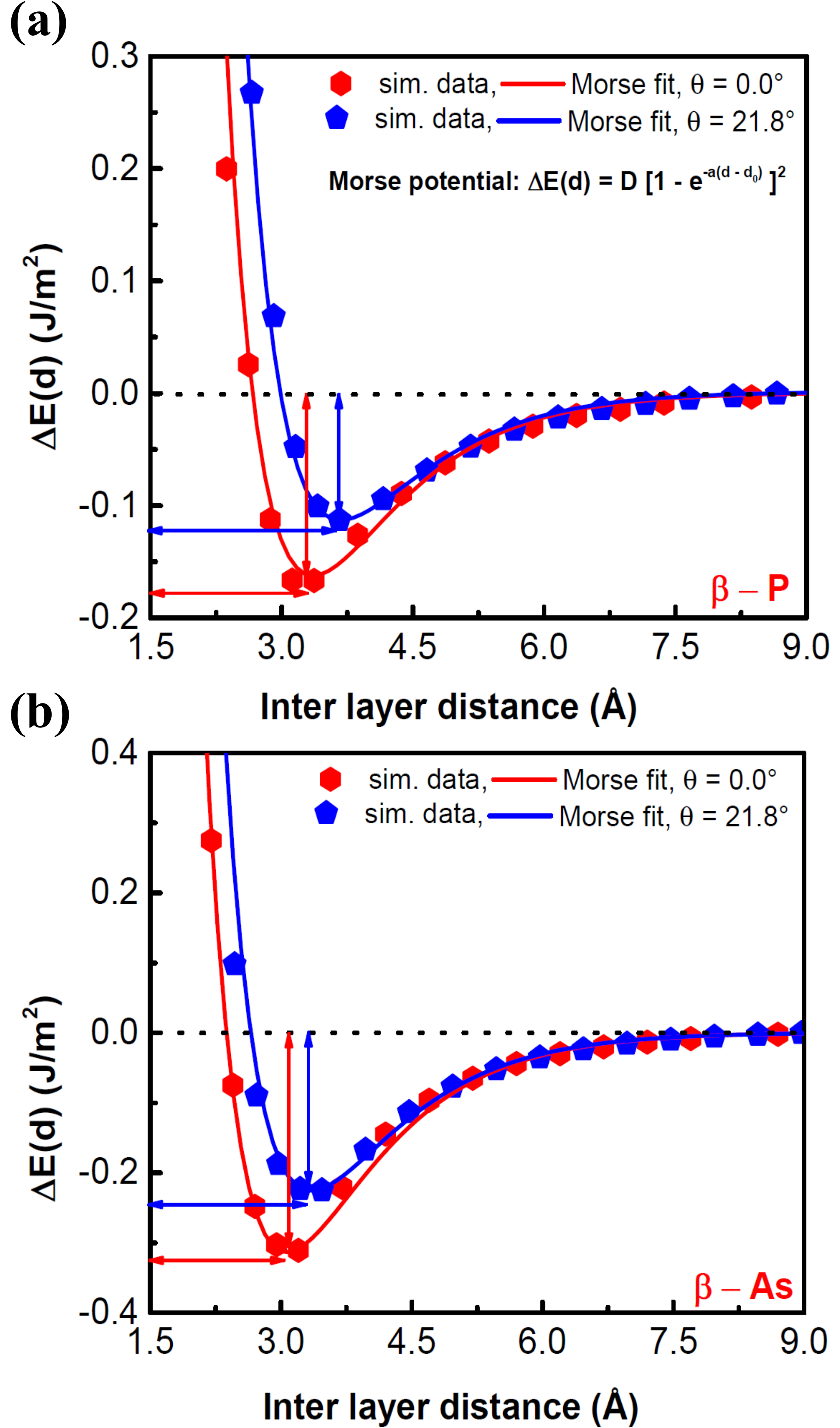}
\caption{Energy of bilayers per unit area (calculated with respect to energy of two ``infinitely'' separated monolayers) plotted as a function of the interlayer separation $d$ for (a) $\beta$-P and (b) $\beta$-As. $\triangle E$ fits well to the Morse potential. The depth of the energy well represents the binding energy. In the case of both the allotropes, binding energy in twisted structures with Moir\'e pattern reduces by $\sim$33\%, resulting in an increase in the interlayer distance of $\sim$10\%.}
\label{fig3}
\end{figure}

The calculated electronic band structures of bilayer AA-stacked $\beta$-P and $\beta$-As are shown in Fig.~\ref{fig1}(b) and Fig.~\ref{fig1}(c), respectively. Evidently, both of them have similar qualitatively features: an indirect bandgap, with conduction band minimum (CBM) and valance band maximum (VBM) located at some point along the $\Gamma$M line. In the valence band, there is another point comparable to that of the VBM energy, lying in between the $\Gamma$K line. Relative weight of the $p_z$ orbitals on different energy bands clearly shows their dominance near the valence band edge. On the other hand, while $p_z$ orbitals have notable contributions near the conduction band edge, $p_x$ and $p_y$ orbitals are found to have equally significant contributions. There is a large anisotropy in the band dispersion, in vicinity of both VBM and CBM. This will lead to highly anisotropic charge carrier effective mass, as reported in case of monolayer allotropes.\cite{Achintya2018,Bajaj} The indirect bandgap values are reported in Table~\ref{table1}, and they are in good agreement with the values reported in the literature.\cite{Barun,kecik1} However, the bandgap in GGA based calculations are known to be underestimated, and nearly two fold increase is reported with use of more accurate hybrid functionals.\cite{Barun,kecik1}

\subsection{Binding energy}

In case of layered materials, monolayers are believed to be stacked on top of each other and  held together by Van der Waals (VDW) interactions. Such forces are isotropic in nature and bonding between two monolayers is expected to be independent of their relative orientation. This hypothesis is tested in case of twisted bilayers of $\beta$-P and $\beta$-As (forming Moir\'e patterns), by comparing their binding energies with that of AA-stacked bilayers. We use the optimized structures of AA-stacked bilayers to generate the Moir\'e superlattices of $\beta$-P and $\beta$-As, by rotating one layer with respect to the other. We select five rotation angles, namely, 10.9$^{\circ}$, 13.2$^{\circ}$, 16.1$^{\circ}$, 21.8$^{\circ}$ and 27.8$^{\circ}$. The Moir\'e superlattices for different relative rotation angles, each of them having a hexagonal unit cell,  are shown in Fig.~\ref{fig2}. Binding between two monolayers is characterized by calculating the difference between the energy of two ``infinitely separated'' monolayers and a bilayer with inter-layer separation $d$:
\begin{equation}
\Delta E_{DFT} (d)=\frac{E_{bilayer} (d) - 2\times E_{monolayer}}{A},
\end{equation}
where $A$ is the area of the unit cell and $d$ is varied over a range from 2.5 to 9 \AA, beyond which $\triangle E$ is found to be independent of $d$. In case of twisted bilayers with Moir\'e patterns, the interlayer distance $d$ is calculated by first measuring the distance between an atom at layer 1 and its closest neighbor at layer 2 and then taking an average of the distances measured between all such pairs. The calculated values of $\Delta E_{DFT} (d)$ for AA stacked and twisted bilayers are shown by the symbols in Fig.~\ref{fig3} (a) and (b) for $\beta$-P and $\beta$-As, respectively. 

\begin{figure*}
\centering
\includegraphics[width=\linewidth]{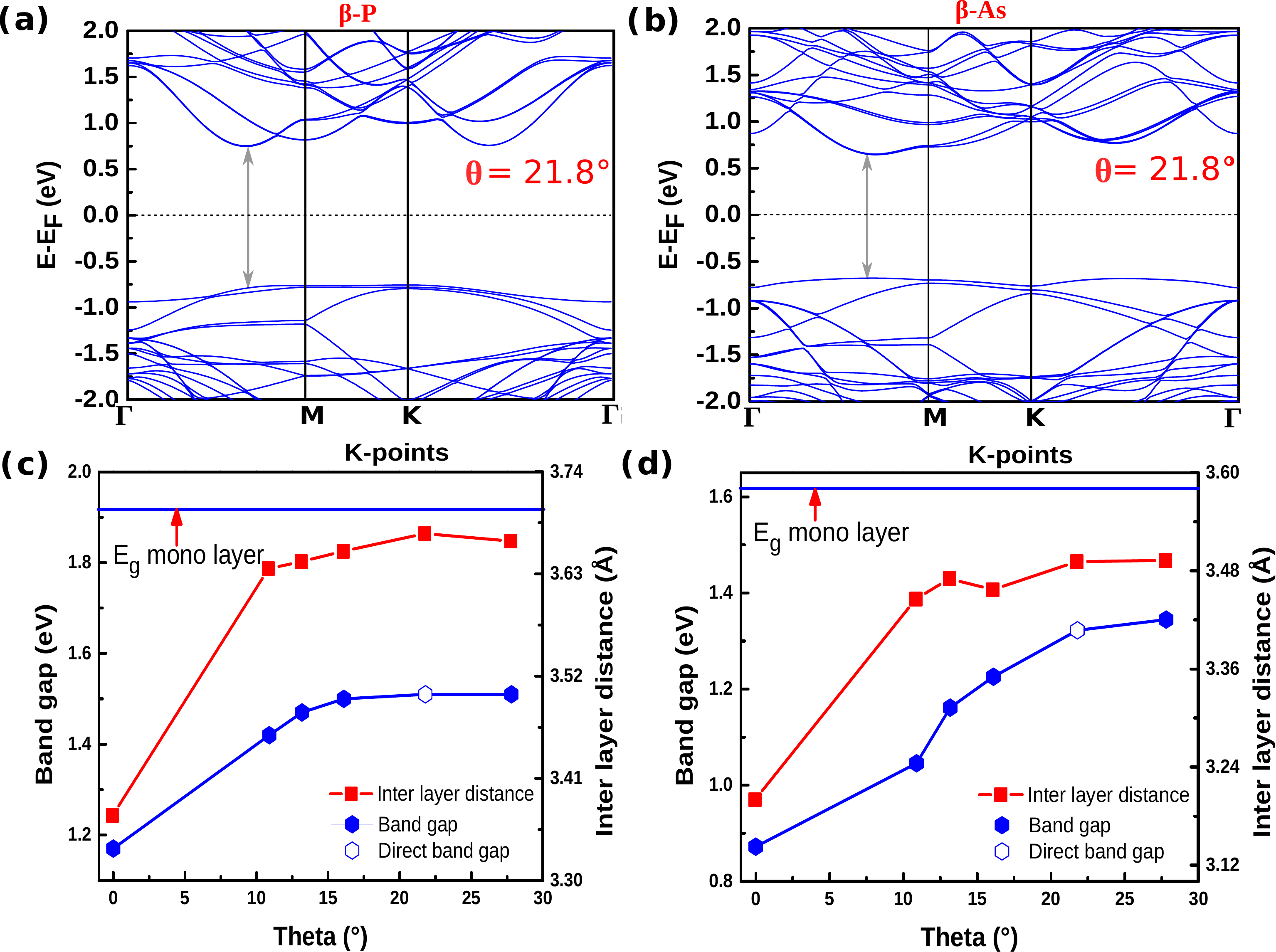}
\caption{Electronic band structure of 21.8$^{\circ}$ rotated Moir\'e pattern of bilayer (a) $\beta$-P and (b) $\beta$-As. For this particular rotation, both the bilayers are converted to a direct bandgap semiconductor, with both the valence and the conduction band edges located in the middle of the $\Gamma$-M line. The variation of the bandgap and the interlayer separation, as a function of the twist angle $\theta$ is shown in (c) $\beta$-P and (d) $\beta$-As. 
Clearly, there is a strong correlation between the bandgap and the interlayer separation of both the Moir\'e superlattices.}
\label{fig4}
\end{figure*}

Further analysis is carried out by fitting the data points obtained from DFT calculations with a Morse potential, which is known to reasonably describe the non-bonded interlayer interactions, particularly at large $d$. The Morse potential is given by $\Delta E_{Morse}(d)= D\left[1-e^{-a(d-d_0)}\right]^2$, where $D$ is the potential depth, $d_0$ is the equilibrium separation and $a$ (taken to be 1 in our case) controls the width of the potential.  The potential depth or $D$, is the difference between the energies at equilibrium separation $d_0$ (equal to 3.35 and 3.2 \AA~for AA-stacked P and As) and at a large value of $d$, where there is no interaction between the monolayers. As shown in Fig.~\ref{fig3}, this energy is almost constant for interlayer distance $\sim$9 \AA, and thus we will use 9 \AA~as the large $d$ value to calculate $D$ for the Morse fit. As shown in Fig.~\ref{fig3} (a)-(b), AA-stacked $\beta$-As has larger (0.30 J/m$^2$) binding energy than that of $\beta$-P (0.16 J/m$^2$). Same is true for the twisted structure with Moir\'e patterns, with $\beta$-As (0.21 J/m$^2$) having nearly twice the binding energy than that of $\beta$-P (0.11 J/m$^2$). Comparing with the AA-stacked bilayers, the binding energy decreases and interlayer separation increases in case of twisted structures. 

The variation of interlayer distance and binding energy as a function of stacking sequence can be attributed to the steric effect: the repulsion between the overlapping electron clouds. Steric effect depends on the size and in-plane interatomic distance of the constituent atoms. For example, larger size and interatomic distance of S atom in MoS$_2$ compared to that of the C atom in graphene, results in a  3 times stronger steric effect in MoS$_2$.\cite{Liu2014} Since P and As atoms are comparable in size with S atom, large steric effect is expected in case of $\beta$-P and $\beta$-As as well. 

Indeed, bilayers of $\beta$-P and $\beta$-As show a wide range of interlayer distances and binding energies for different stacking sequences. The most stable ground-state stacking in the bilayers is the  AA-configuration.\cite{Zhang2015,kecik1} The twisted bilayers with Moir\'e pattern can be considered to be a ``mixture'' of all possible stacking sequences. Thus the binding energy of the twisted structures is expected to be lower than that of AA-stacking. Our calculations reveal nearly 33\% reduction of binding energy and 10\% increase of interlayer separation in 21.8$^\circ$ rotated twisted structures, compared to AA-stacked bilayers. The large dependence of the binding energy on the orientation of the constituent monolayers, clearly highlights the role of steric effects in these bilayers. Thus the interlayer bonding in the bilayers of $\beta$-P and $\beta$-As is not purely of VDW type. Similar observation has also been reported for the other layered allotrope such as black phosphorus.\cite{Shulenburger2015} 

\begin{figure}
\centering
\includegraphics[width=\linewidth]{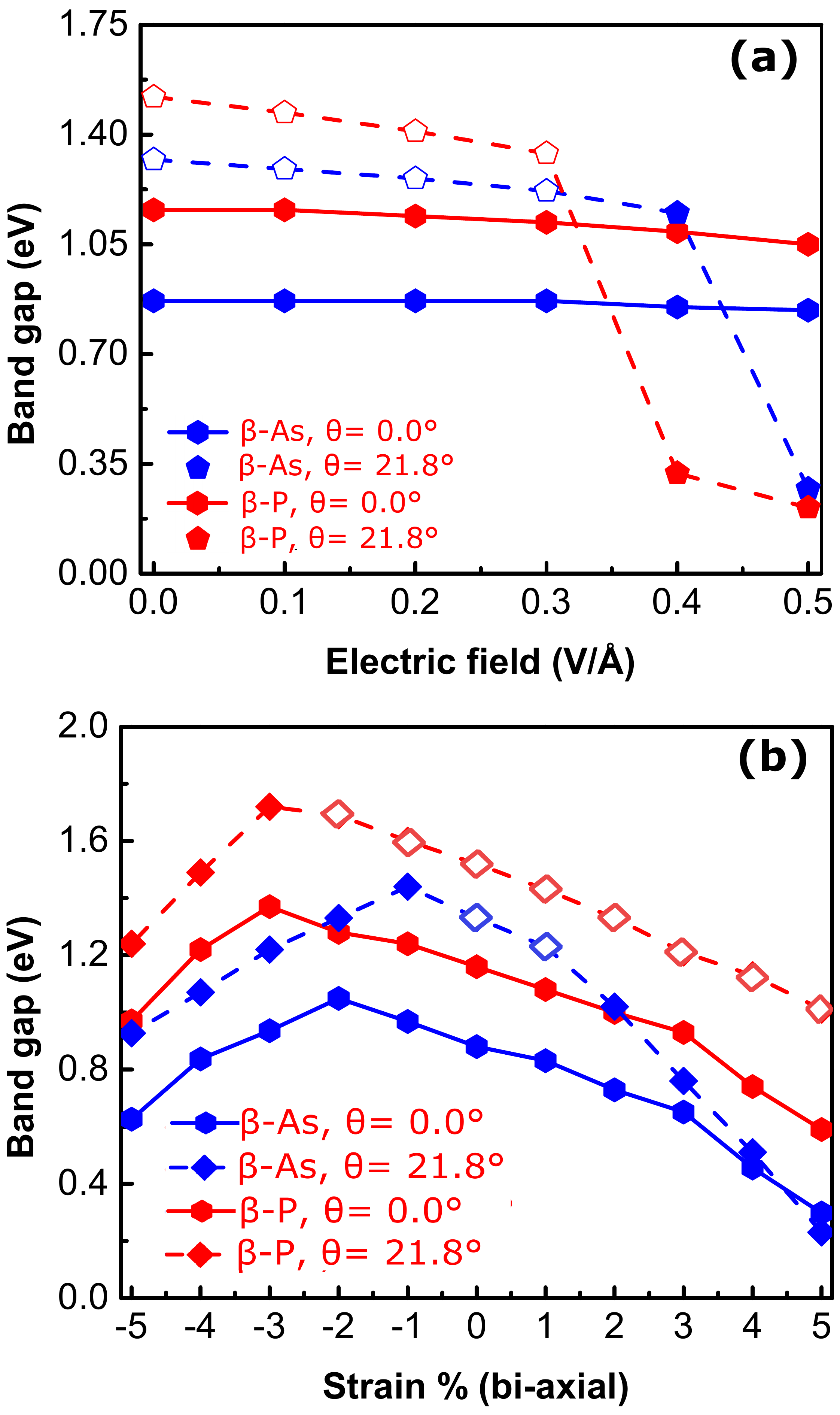}
\caption{(a) The evolution of the bandgap of bilayer $\beta$-P and $\beta$-As with the strength of a vertical electric field. The twisted structures with a larger interlayer separation are impacted more, as expected. (b) Bandgap engineering with bi-axial strain.}
\label{fig5}
\end{figure}

\subsection{Electronic band structure of twisted bilayers}

In the previous subsection, we showed that the formation of the Moir\'e patterns weakens the interlayer interaction. Increased interlayer distance in Moir\'e pattern of $\beta$-P and $\beta$-As leads to the reduction in the overlap among out of the plane $p_z$ orbitals. Since both valence band maximum (VBM) and conduction band minimum (CBM) have significant contributions from the $p_z$ orbitals [see Fig.~\ref{fig1}(b) and (c)], decreasing overlap among such orbitals is expected to have significant effects. 

Two such band structures are shown for $\theta=21.8^\circ$ rotated Moir\'e superlattice of $\beta$-P and $\beta$-As, in Fig.~\ref{fig4}(a) and (b), respectively. Interestingly, for this particular rotation, both the bilayers are transformed to a direct bandgap semiconductor, with the valence and conduction band edges located in the middle of the $\Gamma$-M line. Note that, since the unit cells of the Moir\'e  superlattices are also hexagonal, the high symmetry points in the reciprocal space are the same as in Fig.~\ref{fig1}(b). The magnitude of the bandgap is plotted as a function of rotation angle $\theta$ in Fig.~\ref{fig4} (c) and (d), for $\beta$-P and $\beta$-As, respectively.  As shown in the figures, compared to AA-stacked bilayers, bandgap in Moir\'e superlattices can increase up to $\sim$25\% and 50\% in case of P and As, respectively. While the bandgap values fall within a relatively small window, ranging from 1.4 to 1.5 eV in case of $\beta$-P Moir\'e superlattices, it changes prominently as a function of $\theta$ in case of $\beta$-As, which opens the possibility of bandgap tuning in the latter, by rotating one layer relative to the other. In order to analyze the connection between bandgap magnitude and interlayer separation, we also plot the latter as a function of the rotation angle in Fig.~\ref{fig4} (c) and (d). Evidently, the bandgap enhancement is highly correlated with the increasing interlayer separation and the decreasing overlap among the $p_z$ orbitals of adjacent layers. 

Other than the $\theta=21.8^\circ$, rest of the Moir\'e superlattices are indirect bandgap semiconductors as shown in Fig.~\ref{fig4}(c) and (d). However, due to the flatness of the valence band, the difference between the direct and indirect bandgap is very small in most of the cases [see Fig.~S1 and S2 in the Supporting Information]. The flatness of the bands in vicinity of the band edges also gives rise to a large density of states, which is conducive to magnetic and other many-body instabilities. A recent example is that of the `magic angle' twisted bilayer graphene, which shows superconducting instability.\cite{Santos, Andrei, Mallet, Mengxi} 

\subsection{Band structure modulation with electric field} 
We now explore the possibility of bandgap engineering in these twisted bilayers by means of a vertical electric field. In case of $\beta$-P and $\beta$-As, very large electric field (about 0.5 V/\AA) is known to reduce the bandgap, ultimately leading to an insulator to metal transition around 0.7 to 1 V/\AA.\cite{Barun,Mardanya} Moreover, an interesting topological transformation is also reported for monolayer $\beta$-As, at nearly 1 V/\AA~ electric field.\cite{Mardanya} 

A similar trend is observed in case of Moir\'e superlattices, where bandgap magnitudes decrease sharply beyond 0.3 [0.4 V/\AA] for bilayer $\beta$-P [$\beta$-As]. A comparison of bandgap modification (with increasing electric field strength) in AA-stacked bilayers and Moir\'e super-lattices is shown in Fig.~\ref{fig5} (a). Interestingly, the latter (originally a direct bandgap semiconductor) is found to be converted to an indirect bandgap semiconductor at high electric electric field, both in case of $\beta$-P  and $\beta$-As. This transition happens because the conduction band edge is shifted to the $\Gamma$ point (see Fig.~S3 and S4 in SI). The computationally predicted electric field values of 0.3-0.4 V/\AA~ can further be reduced if we account for the presence of ripples and other structural imperfections.\cite{Agnihotri} 

\subsection{Band structure modulation with strain}
Monolayers of $\beta$-P and $\beta$-As also show significant bandgap modulation under the influence of external strain.\cite{David,Mardanya,kecik1} A similar trend is observed in case of AA-stacked bilayers and Moir\'e superlattices of $\beta$-P and $\beta$-As. As shown in Fig.~\ref{fig5}(b), the bandgap decreases monotonically under tensile strain. On the other hand, under compression, bandgap initially increases, followed by a decrease at higher value of strain. Evolution of electronic band structures as a function of strain for AA-stacked $\beta$-P and $\beta$-As are shown in SI (see Fig.~S5 and S6). We find that, not only the magnitude of bandgap decreases under strain, band edges also shift from their original location. For example, in case of AA-stacked $\beta$-P, CBM and VBM shifts to the K-point and $\Gamma$-point at 2\% and 4\% compressive strain, respectively (see Fig.~S5 in SI). On the other hand, in case of AA-stacked $\beta$-As, VBM moves to the $\Gamma$-point under 3\% compressive strain, while CBM shifts to the $\Gamma$-point at 4\% tensile strain (see Fig.~S6 in SI). However, we do not find any indirect to direct bandgap conversion within $\pm$5\% strain in case of AA-stacked $\beta$-P and $\beta$-As.

Interestingly, for  Moir\'e superlattices, the bilayers can also convert from a direct to an indirect bandgap semiconductor under strain. This is more prominent in case of $\beta$-As Moir\'e superlattice ($\theta=21.8^\circ$), which changes to an indirect bandgap semiconductor under relatively small tensile (2\%), as well as compressive (1\%) strain. On the other hand, $\beta$-P Moir\'e superlattice ($\theta=21.8^\circ$) undergoes such a transition only under compression (3\%). A detailed analysis reveals that the band edges shift to different valleys as a function of strain, leading to a direct to indirect bandgap transformation. For example, in case of  $\beta$-P Moir\'e superlattice, the valence band edge moves to the $\Gamma$ point and the conduction band edge moves to the K-point at 3\% compression (see Fig.~S7 in SI). On the other hand, in case of $\beta$-As Moir\'e superlattice, the valence band edge moves to the $\Gamma$ point at 1\% compressive strain, while a 2\% tensile strain triggers a similar shift to the conduction band edge. (see Fig.~S8 in SI).

\begin{figure}
\centering
\includegraphics[width=\linewidth]{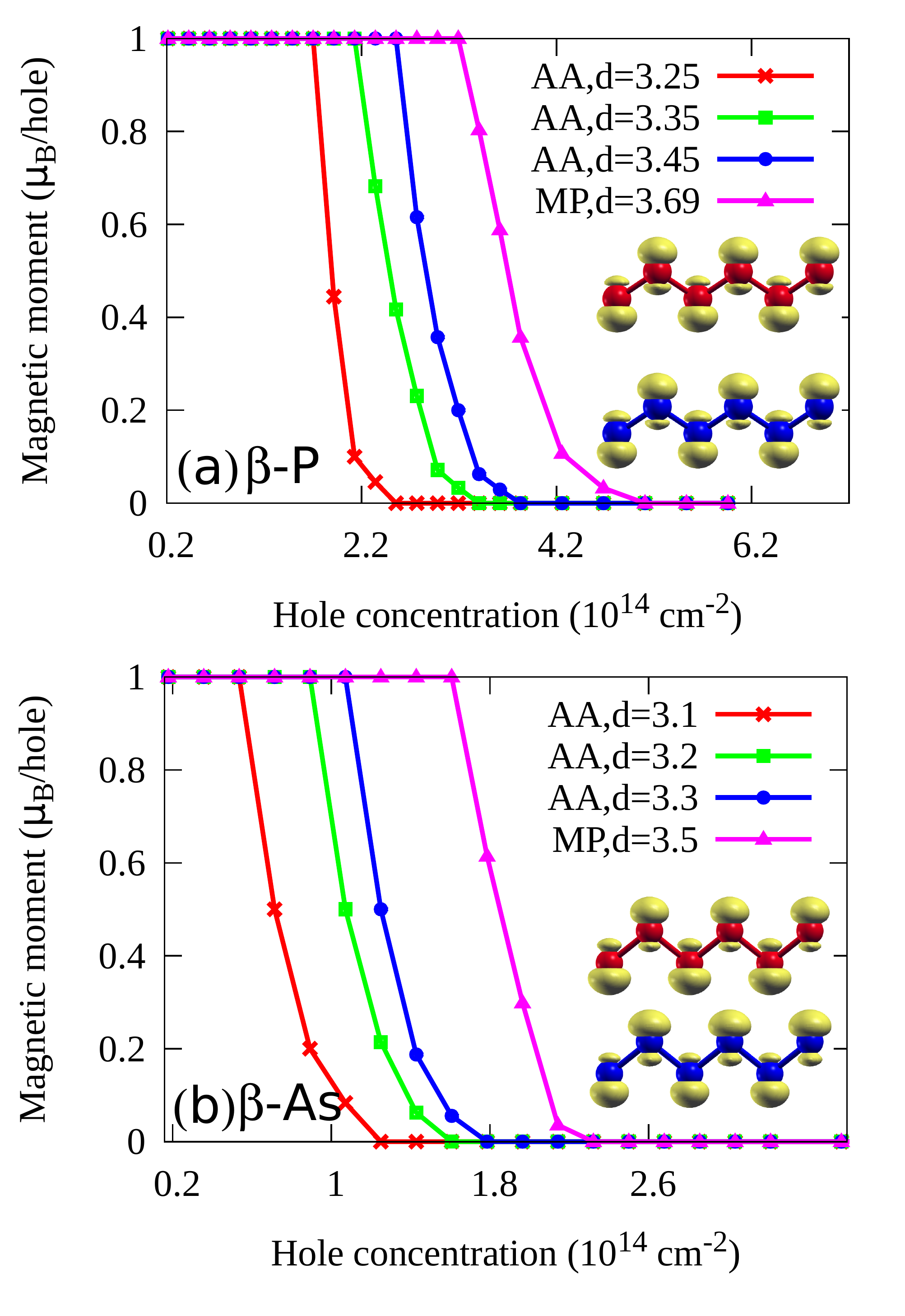}
\caption{Variation of magnetic moment as a function of hole concentration for AA-stacked bilayer (a) $\beta$-P and (b) $\beta$-As. The critical value of hole doping, up to which ferromagnetism exists, increases with interlayer distance $d$. The inset in both the panels, show the out of plane spin density. Clearly the major spin contribution is from  the $p_z$ orbitals of P and As atoms, whose interaction is $d$ dependent. In case of Moir\'e super-lattice ($\theta=21.8^{\circ}$) with larger $d$, the interlayer decoupling supports the ferromagnetic ground state up to higher doping concentration.}
\label{fig6}
\end{figure}

\subsection{Magnetic instability} 
As shown Fig.~\ref{fig1}(b)-(c) and Fig.~\ref{fig4}(a)-(b), the valence band of bilayer $\beta$-P and $\beta$-As have a relatively flat shape near the VBM. This results in a van Hove singularity in the electronic density of states near the valence band edge, which can be explored via hole doping. The van Hove singularity is conducive to quantum many body instabilities such as magnetism. In fact, magnetism  arising from the van Hove singularity induced Stoner instability in monolayers and bilayers of P and As has already been predicted over a  over a broad range of hole doping.\cite{botao2017,zhou2017} 

Here we explore the existence of a ferromagnetic instability in hole doped bilayers of twisted $\beta$-P and $\beta$-As. The calculated magnetic moment (per hole) in the Ferromagnetic ground state is shown in Fig.~\ref{fig6}. Our calculations reveal that the magnetic moment is equal to the number of holes in the system up to certain value of doping, beyond which both the bilayers fall back to the non-magnetic state rapidly. The critical value of hole concentration at which the crossover takes place is nearly double in AA-stacked $\beta$-P, than that of $\beta$-As. This is in good agreement with the reported data in the literature.\cite{botao2017,zhou2017}  Note that, prediction of a ferromagnetic ground state up to $10^{14}$ cm$^{-2}$ hole concentration is very encouraging, because it is within the range of current experimental capabilities, as demonstrated for gate-tunable MoS$_2$ device.\cite{Ye2012}  

As shown in the respective insets for Fig.~\ref{fig6}, the spin density is predominantly localized in the $p_z$ orbitals of the P and As atoms. Thus, the increase in the interaction of the $p_z$ orbitals with decreasing interlayer separation in bilayers is likely to result in the weakening of the magnetic ground state. This is confirmed by calculating the magnetic moment by changing the interlayer separation $d$ (slightly above and below the equilibrium value) for AA-stacked bilayers in Fig.~\ref{fig6}. As expected we find that 
the critical value of hole doping, up to which the AA-stacked bilayers remain in ferromagnetic ground state, gradually decreases as the monolayers are brought closer to each other. This is also reflected in the fact that  the ferromagnetic ground state survives up to higher hole concentration in case of Moir\'e superlattices (see Fig.~\ref{fig6} for results for $\theta=21.8^\circ$), which has a larger interlayer distance and smaller interactions among the $p_z$ orbitals. In fact, the ferromagnetic ground state can exist up to nearly 50\% higher hole doping in Moir\'e superlattices as compared to the AA-stacked bilayers, making the former more suitable for magnetic and spintronic applications.

\section{Conclusion}
\label{sec4}
In conclusion, we have investigated the evolution of interlayer coupling in twisted $\beta$-P and $\beta$-As bilayers. Our calculations reveal nearly 33\% reduction in binding energy, leading to 10\% increase of interlayer separation in Moir\'e superlattices of $\beta$-P and $\beta$-As, compared to AA-stacked bilayers. We show that the band edges (particularly the VBM) are dominated by the out of plane $p_z$ orbitals in both $\beta$-P and $\beta$-As. Thus, increasing the interlayer separation reduces the overlap among $p_z$ orbitals from adjacent planes. This interlayer decoupling in Moir\'e superlattices leads to significant changes in the electronic, as well as magnetic (with hole doping) properties of $\beta$-P and $\beta$-As. 

We have shown that compared to AA-stacked bilayers, bandgap increases by 25-50\% in Moir\'e superlattices. Furthermore, on account of a relatively flat valence band, the difference between the direct and indirect bandgap is very small. In some cases, twisted bilayers are found to be direct bandgap semiconductors, while the AA-stacked bilayers are indirect bandgap semiconductors. Similar to other 2D materials, bandgap can be tuned by applying strain, as well as an electric field normal to the plane of the twisted bilayers. 

Due to the Stoner instability originating from the van Hove singularity of density of states near the valence band edge, a ferromagnetic ground state is observed in bilayer $\beta$-P and $\beta$-As. The critical value of hole concentration, up to which the ferromagnetic ground state is sustained, increases by $\sim$50\% in Moir\'e superlattices, compared to AA-stacked bilayers. Thus, twisted bilayers are better candidates than AA-stacked $\beta$-P and $\beta$-As for the purpose of magnetic and spintronic applications. 


\section{Acknowledgments}
The authors acknowledge funding from the Ramanujan fellowship research grant, the DST Nanomission project and SERB (EMR/2017/004970). The authors also thank the computer center of IIT Kanpur for providing HPC facility.
\bibliography{ref}
\end{document}